\documentclass[prb,aps,twocolumn,superscriptaddress,showpacs,10pt]{revtex4-2}
\usepackage{graphicx,amsfonts,times,bm,amsmath,verbatim,color,array}
\usepackage{lineno}
\usepackage{xcolor}
\usepackage{algorithmic}
\usepackage{braket}
\usepackage[colorlinks,urlcolor=blue,citecolor=blue,linkcolor=blue]{hyperref}

\begin{document}

\title{Approximate half-integer quantization in anomalous planar transport in $d$-wave altermagnets}
\author{Srimayi Korrapati}
\affiliation{Department of Physics and Astronomy, Clemson University, Clemson, SC 29634, USA}
\author{Snehasish Nandy}
\affiliation{Department of Physics, National Institute of Technology Silchar, Assam, 788010, India}
\author{Sumanta Tewari}
\affiliation{Department of Physics and Astronomy, Clemson University, Clemson, SC 29634, USA}

\begin{abstract}
 We investigate anomalous planar transport phenomena in a recently identified class of collinear magnetic materials known as $d$-wave altermagnets. The anomalous planar effects manifest in a configuration when the applied electric field/temperature gradient, magnetic field, and the measured Hall voltage are all co-planar, but the planar magnetic field is instrumental in breaking $\hat{C}_{4z}\hat{\mathcal{T}}$ symmetry of the $d$-wave altermagnet, where $\hat{\mathcal{T}}$ is the time reversal operator, resulting in a Zeeman gap at a shifted Dirac node and a nonzero Berry curvature monopole. We demonstrate that these systems exhibit nearly half-quantized anomalous planar Hall and planar thermal Hall effects at low temperatures that persist over a range of magnetic fields.  The angular dependence of the planar transport reveals a $\cos2\phi$ dependence on the magnetic field direction, where $\phi$ is the azimuthal angle made by the magnetic field. We also discuss the anomalous planar Nernst effect, or transverse thermopower, and demonstrate that the Nernst conductivity peaks when the chemical potential lies just outside the induced Zeeman gap and vanishes within the gap. We further explore the dependence of all three coefficients on the polar and the azimuthal angle of the magnetic field when it is rotated in the full $3D$ space. Our results reveal the presence of approximately \textit{half}-quantized anomalous planar thermal Hall plateau for a range of in-plane magnetic fields without requiring topological superconductivity and conducting Majorana modes, and can be probed in experiments in $d$-wave altermagnets. 
\end{abstract}
\maketitle
\section{INTRODUCTION}\label{seclevel1}
\begin{figure*}[htb]
\centering
\includegraphics[width=0.95\textwidth]{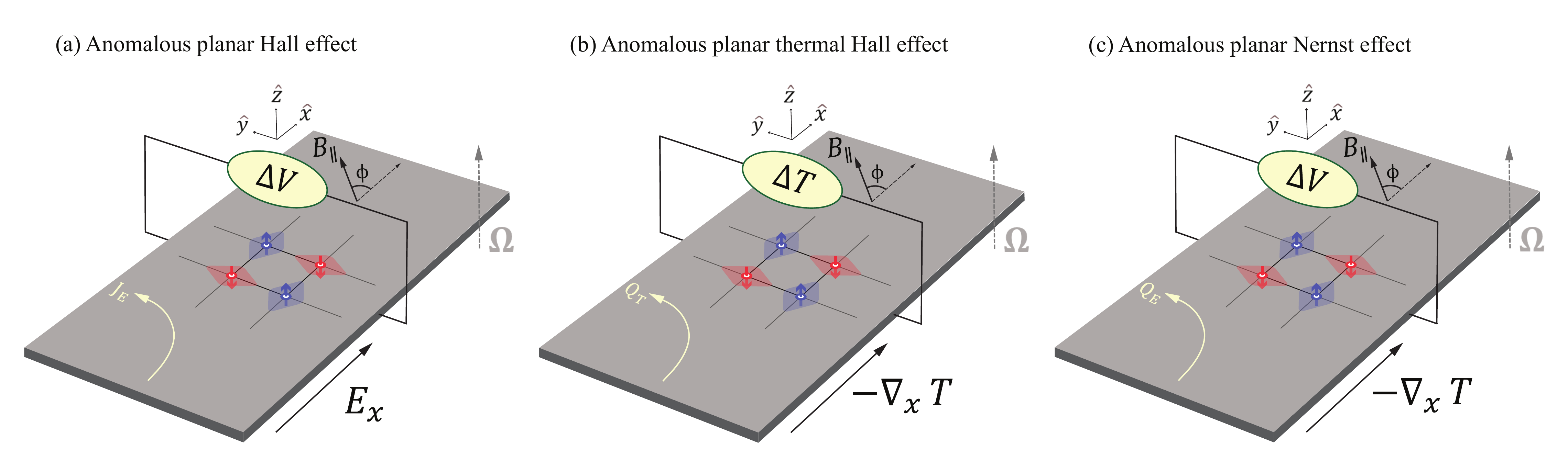}
\caption{Schematic diagram of the experimental setup for measuring anomalous planar Hall, planar thermal Hall and planar Nernst effects in altermagnets. The magnetic ordering on the square lattice with $\hat{\mathcal{C}_{4z}}\hat{\mathcal{T}}$ symmetry is indicated by the spin-up sublattice in blue and spin-down sublattice in red. \textcolor{black}{Here, the N\'{e}el vector is along the out-of-plane $\hat{z}$ direction}. The spin-up and spin-down sublattices are connected via a $\pi/2$ rotation about the center. An in-plane magnetic field $\bm{B_{\parallel}}$ modifies the Berry curvature ($\bm{\Omega} = \Omega_z$) to produce a non-zero Berry curvature Monopole (see Fig.~\ref{fig:hamiltonian}) under the presence of which (a) a longitudinal electric field induces a transverse potential difference ($\Delta V$) in the Hall effect, (b) a longitudinal temperature gradient ($-\bm{\nabla} T$) gives rise to a transverse thermal gradient ($\Delta T$) in the thermal Hall effect and (c) a longitudinal temperature gradient gives rise to a transverse potential difference ($\Delta V$) in the Nernst effect.}
\label{fig:exp}
\end{figure*}

A recently discovered subclass of magnetic materials, known as altermagnets (AMs), has opened a new frontier in condensed matter physics by exhibiting a novel form of collinear magnetic order that transcends the conventional dichotomy of ferromagnetism and antiferromagnetism~\cite{Kusunose_2019, Zunger_2020, Libor_2021, Jungwirth_2022, Tomas_2022}. Altermagnets are uniquely characterized by two key features: (i) the absence of net magnetization despite the breaking of time-reversal symmetry (TRS), and (ii) momentum-dependent spin splitting in their electronic bands in the absence of spin-orbit coupling~\cite{Jungwirth_2022, Tomas_2022, Igor_2022, Sinova_2022, Spaldin_2024}. The vanishing net magnetization arises from an alternating arrangement of magnetic moments in both real and momentum space, distinguishing AMs from ferromagnets, which also break TRS but possess a net magnetization. In contrast to conventional antiferromagnets, where magnetic sublattices are related through inversion or translation, resulting in symmetry-protected spin-degenerate bands, AM sublattices are connected via proper or improper rotations and/or reflections (see Fig.~\ref{fig:exp}), leading to non-relativistic spin splitting without relying on spin-orbit coupling~\cite{Blundell_2001, Jungwirth_2022, Tomas_2022, Spaldin_2024}. These distinct symmetry and electronic features underpin a wide array of novel physical phenomena~\cite{Song_2022, Kriegner_2023, Bai_2023, Tsymbal_2023, Yan_2023, Jacob_2023, Yugui_2024, Zhang_2024, Papaj_2023, Ghorashi_2023, Cheng_2023, Jeffrey_2024, Hariki_2024, Antonenko_2024, Samokhin_2024, Nandy_2025}, with significant implications for fundamental physics and technological applications, including superconducting phenomena \cite{fukaya2025superconducting}, spintronics \cite{fu2025allelectricallycontrolledspintronicsaltermagnetic} and quantum information processing~\cite{Yao_2024}.

\begin{figure*}[!ht]
\centering
\includegraphics[width=0.98\textwidth]{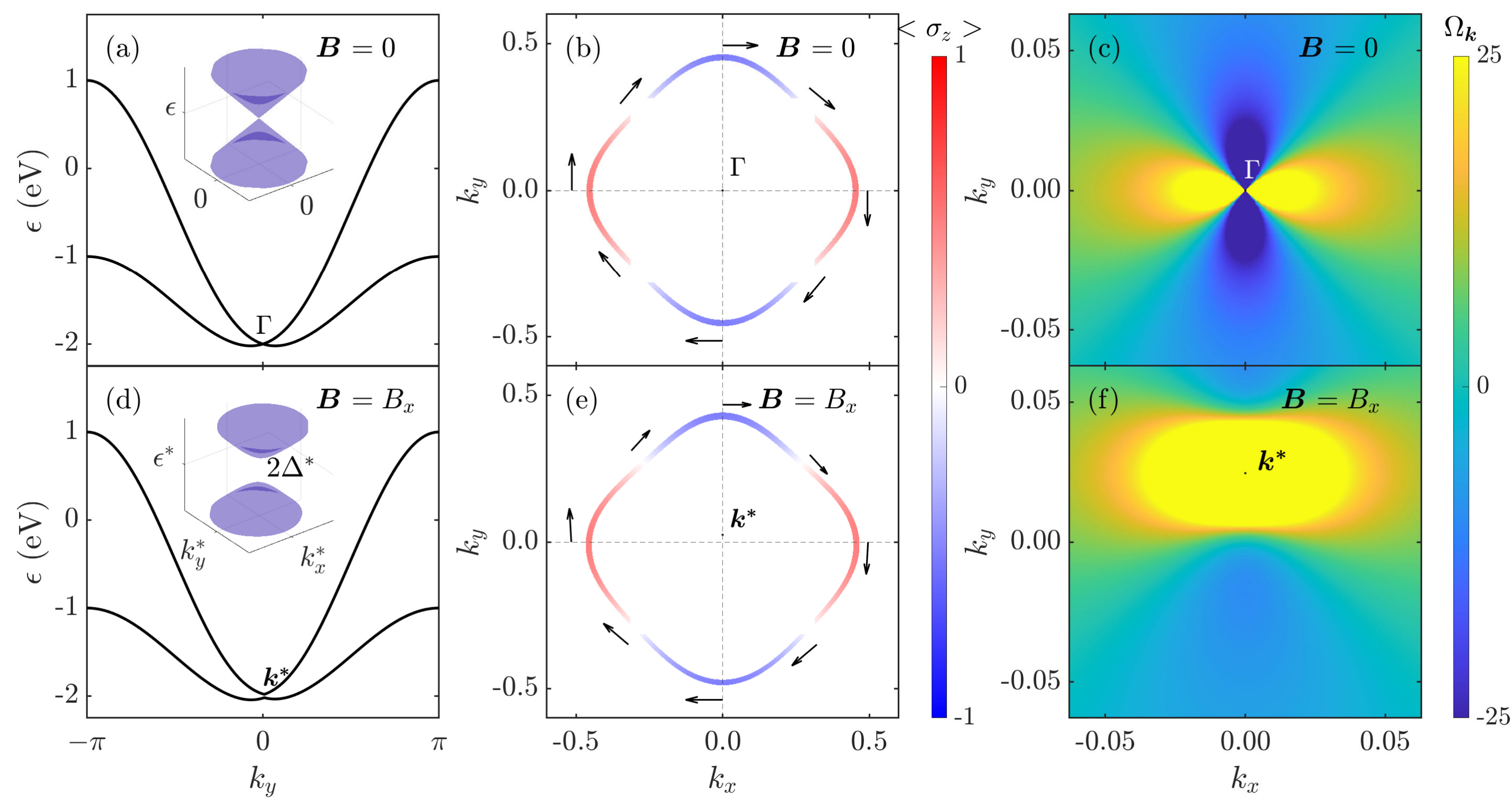}
\caption{
Panel (a) shows the energy dispersion of the Hamiltonian (\ref{eq:hamLat}) in the absence of an applied magnetic field with the inset showing the gapless Dirac cone at the $\Gamma$ point. In panel (d), an applied magnetic field $B_x$ shifts the Dirac point from $\Gamma$-point to $\bm{k}^*$ and opens up a gap of $2|\Delta^*|$ (see Eq.~(\ref{eq:gap})). Panels (b) and (e) show the spin textures of the single anisotropic spin-split Fermi surfaces when the chemical potential is set at the band-touching point in the absence of $\bm{B_\parallel}$ and at center of the Dirac energy gap in the presence of $\bm{B_\parallel} = (7\text{ T},0)$, respectively. The black arrows indicate in-plane spin-polarization while the red-blue color bar represents the out-of-plane spin-polarization. Panel (c) shows the momentum-resolved Berry curvature distribution of the lower band having a quadrupole nature in the absence of a magnetic field (where the positive and negative lobes of the distribution exactly cancel each other out, rendering the BCM zero) while panel (f) reveals the effective monopole characteristic induced by an in-plane magnetic field $B_x$ (where there is no longer an equal magnitude of positive and negative regions, yielding a finite BCM). The parameters used here are $t=0.5$ eV, $\lambda =0.1$ eV, $t_{\text{AM}} =0.25 $ eV, comparable to parameters used in Refs. \cite{Rao_2024,Yan_2023}.}
\label{fig:hamiltonian}
\end{figure*}

The family of Hall effects, manifested as a transverse voltage in response to an applied current in metals or semiconductors, has been instrumental in uncovering topological phases of matter and enabling numerous applications~\cite{Karplus_1954, Klitzing_1980, Haldane_1988, Kane05, Nagaosa_2010, Liu_2016, Das_2021}. Among them, the intrinsic anomalous Hall effect (AHE)~\cite{Nagaosa_2010, Ado_2015}, which arises in the linear-response regime without an external magnetic field, has garnered significant theoretical and experimental interest. This is primarily because the linear AHE directly probes the Berry curvature (${\Omega_{\bm{k}}}$), a central concept in modern topological band theory derived from the electron wavefunction \cite{xiao10}. However,  the linear anomalous Hall effect, governed by the Berry curvature monopole (BCM), quantified as the integral of the Berry curvature over the Brillouin zone over occupied electronic states \cite{xiao10}, is non-zero only in systems with broken time-reversal symmetry, as dictated by the Onsager reciprocity relations \cite{Landau_1980}. In contrast, TRS-invariant systems, where the linear AHE vanishes, can exhibit a second-order (nonlinear) anomalous Hall effect governed by the Berry curvature dipole, defined as the integral of the first-order moment of the Berry curvature ($\partial k_{a}\Omega_{\bm{k}}$) over the occupied states~\cite{Sodemann_2015, Sodemann_2019, Nandy_21}. Recent work has demonstrated that $d$-wave AMs, which preserve the combined fourfold rotation ($\hat{C}_{4z}$) and time-reversal ($\hat{\mathcal{T}}$) symmetries, can host a third-order anomalous Hall effect as the leading-order response~\cite{Fang_2024}. In particular, despite the individual breaking of both the $\hat{C}_{4z}$ and $\hat{\mathcal{T}}$ symmetries, the BCM and Berry curvature dipole vanish under the combined $\hat{C}_{4z}\hat{\mathcal{T}}$ symmetry which enforces the transformation $\Omega(k_x,k_y)\rightarrow - \Omega(-k_y,k_x)$. The Berry curvature monopole vanishes because in the 2D Brillouin zone integral, $\Omega(k_x,k_y)$ is exactly canceled by $\Omega(-k_y,k_x)$ (see Fig.~\ref{fig:hamiltonian}(c)). Furthermore, the derivatives transform as $\partial k_a \Omega(k_x,k_y)\rightarrow -\partial k_a \Omega(-k_x,-k_y)$, making the first moment of Berry curvature a function that is odd in $k_a$. The Berry curvature dipole, which then involves the integral of an odd function in momentum over the Brillouin zone,  vanishes identically. Consequently, a finite third-order Hall response emerges as the first non-trivial Hall response and is driven by the Berry curvature quadrupole defined as the integral of the second-order moment of the Berry curvature ($\partial k_{a}\partial k_{b} \Omega_{\bm{k}}$) over the occupied states~\cite{Fang_2024,Zhang_2023, PhysRevB.110.195119}. In addition, recent studies have identified both linear and nonlinear Hall currents in altermagnetic systems arising from various mechanisms, including the wavepacket magnetic moment~\cite{Ganesh_2025}, the interplay of incident light and strain~\cite{Farajollahpour_2025, Keigo_2025}, domain-wall-induced orbital magnetization~\cite{Sorn_2025}, and the application of external magnetic fields~\cite{Rao_2024, Antonenko_2024}. Recently, a half-quantized Hall response has been shown to exist in systems with hexagonal warping \cite{PhysRevB.106.L241105}, in altermagnets in proximity to weak topological insulators \cite{PhysRevB.111.045407, PhysRevB.111.045409}, and in $d_{xy}$-wave altermagnets \cite{PhysRevB.109.L180411}.

In this work, we study the anomalous transport phenomena in two-dimensional (2D) $d_{x^2-y^2}$-wave altermagnets (such as $\text{Ru}\text{O}_2$ \cite{Ghorashi_2023}, $\text{Mn}_5\text{Si}_3$ \cite{PhysRevB.110.L220411}) with substrate-induced Rashba spin-orbit coupling (RSOC) in the presence of an in-plane magnetic field. We show that this system exhibits anomalous planar Hall, thermal Hall, and Nernst effects, mediated by a non-trivial Berry curvature monopole induced by the planar magnetic field, which breaks the combined $\hat{C}_{4z}\hat{\mathcal{T}}$ symmetry. Interestingly, when the chemical potential lies within the gap at the shifted Dirac node induced by the in-plane magnetic field, nearly half-quantized anomalous planar Hall and anomalous planar thermal Hall conductivities emerge. This approximate quantization originates because, in the presence of an in-plane magnetic field, RSOC opens a gap at the Dirac point, leading to a \textcolor{black}{BCM} $\nu = \pm 1/2$, where $\nu$ is defined as the integral of the Berry curvature over the Brillouin zone. When the chemical potential lies within the Dirac gap, the magnitude of which is controlled by the magnetic field, only one of the bands is occupied, and the system evinces approximately half-quantized anomalous planar Hall and planar thermal Hall effects. We show that the approximately half-quantized anomalous planar thermal Hall coefficient forms a plateau as a function of the in-plane magnetic field, as shown in Fig.~3(c). Such a half-integer quantized anomalous thermal Hall plateau has been associated with non-Abelian states with conducting Majorana modes \cite{Kasahara2018,Yokoi2021,PhysRevLett.121.026801}, but in this paper, we demonstrate that this exotic effect can arise even from a simple Berry curvature effect arising at a Dirac gap induced by a planar magnetic field.  

The approximate quantization is gradually suppressed either by raising the temperature, causing thermal broadening of the Fermi–Dirac distribution and activating contributions from both bands with opposite Berry curvature, or by tuning the magnetic field such that the chemical potential is outside the Dirac gap. With rotating the magnetic field orientation, we study angular dependencies of the planar transport responses and find that these responses are even in the in-plane magnetic field ($\bm{B}_{\parallel}$) with a period of $\pi$ emerging from the underlying $\cos2\phi$ dependence of the energy gap, where $\phi$ is the azimuthal angle between the magnetic and the electric field (thermal gradient). 
Moreover, we find that the system encounters topological phase transitions as the in-plane magnetic field is rotated within the plane characterized by a change in the \textcolor{black}{BCM} between $\nu = + 1/2$ and $\nu = -1/2$ as a function of the azimuthal angle $\phi$.  
We further study the planar Nernst response in the same configuration and find that the peak in the planar Nernst, also called transverse thermopower, occurs when the chemical potential lies just outside  the gap induced at the shifted Dirac node by the in-plane magnetic field and vanishes when the chemical potential is within the gap. With increasing temperature, the chemical potential where the peaks occur moves away from the induced gap. The sign of these peaks changes with the same $\cos2\phi$ dependence on the in-plane magnetic field angle as the Dirac mass term, the topological phase transitions, and the nearly half-quantized Hall and thermal Hall responses.
We also study these responses as the magnetic field is rotated throughout the full $3D$ space, as a function of the polar angle $\theta$ and the azimuthal angle $\phi$. We further identify the temperature regions where the Wiedemann-Franz law and the Mott relation hold, and show that their validity breaks down with increasing temperatures.  

The remainder of the paper is organized as follows: In Sec. \ref{sec: formalism}, we present the general expressions of anomalous planar Hall, thermal Hall and Nernst effects within the semiclassical Boltzmann formalism. In Sec. \ref{sec: model}, we introduce the model Hamiltonian for a 2D $d$-wave AM and analyze its energy dispersion, spin-split Fermi surfaces, and Berry curvature distributions both with and without an applied in-plane magnetic field. In Sec. \ref{sec: results}, using the general expressions of anomalous planar responses introduced in Sec. \ref{sec: formalism}, we numerically compute the anomalous planar Hall, thermal Hall, and Nernst conductivities for the $d$-wave AM. We examine the dependence of these transport coefficients on magnetic field strength and orientation, chemical potential, and temperature. We further investigate the validity of Wiedemann-Franz law and Mott relation in this system. Finally, we end with
a brief conclusion in Sec. \ref{sec: conclusions}.

\section{\label{sec: formalism}Planar Anomalous Transport}
In the presence of external perturbative fields—namely, an electric field $\bm{E}$ and a temperature gradient $-\bm{\nabla}T$—the linear response of the system yields the charge current $\bm{J}$ and thermal current $\bm{Q}$ in the following form:
\begin{equation}\label{eq:currents}
    \begin{split}
        J_{a} &= \sigma_{ab}E_{b} + \alpha_{ab}(-\nabla_{b} T),\\
        Q_{a} &= \frac{\alpha_{ab}}{T}E_{b}+\kappa_{ab}(-\nabla_b T),
    \end{split}
\end{equation}
where $a$ and $b$ denote spatial indices ($x$, $y$, $z$), and the tensors $\sigma_{ab}$, $\alpha_{ab}$, and $\kappa_{ab}$ correspond to the charge conductivity, thermoelectric, and thermal conductivity tensors, respectively. In this work, we focus on the evaluation of transverse transport coefficients induced by an applied temperature gradient in the presence of an in-plane external magnetic field, as shown in Fig.~\ref{fig:exp}.

Within the semiclassical Boltzmann framework under the relaxation time approximation, the anomalous Hall conductivity due to Berry curvature contributions can be expressed as~\cite{Nagaosa_2010}:
\begin{equation}\label{eq:AHE}
\sigma_{xy} = \frac{e^2}{\hbar} \int _{\bm{k}} f^{n}_{\bm{k}} \Omega^n_{\bm{k}},
\end{equation}
where the integration is over the first Brillouin zone and $\int_{\bm{k}} \equiv \sum_{n} \int d^{d}k/(2\pi)^{d}$ with $d$ being the dimensionality and $n$ being the band index; $\Omega^n_{\bm{k}}=-2\text{Im} \braket{  \frac{\partial u_{n\bm{k}}}{\partial k_x}|\frac{\partial u_{n\bm{k}}}{\partial k_y}}$ is the Berry curvature for the state $\ket{u_{n\bm{k}}}$, $f^n_{\bm{k}} \equiv f(\epsilon^{n}_{\bm{k}}-\mu)$ is the equilibrium Fermi-Dirac distribution function and $\mu$ is the chemical potential. 

It is worth emphasizing that the intrinsic anomalous Hall effect is typically defined as the integral of the Berry curvature over the occupied states at $\bm{B}=0$, and it yields a finite contribution for systems with broken TRS \cite{xiao10}. However, in the altermagnetic system under consideration with $\hat{\mathcal{C}_{4z}}\hat{\mathcal{T}}$ symmetry, the non-trivial Berry curvature integrates to zero and no intrinsic anomalous Hall effect is expected at zero magnetic field \cite{Zou2024}. Remarkably, the introduction of a planar magnetic field via the Zeeman coupling explicitly breaks the $\hat{\mathcal{C}_{4z}}\hat{\mathcal{T}}$, thereby inducing a finite Berry curvature monopole. This gives rise to a transverse Hall voltage in response to a coplanar electric field, which is purely topological in origin and distinct from classical mechanisms. Moreover, when the magnetic field is applied along the $z$-direction, the Zeeman effect can coexist with Lorentz force effects or Landau quantization, making their contributions hard to separate. In contrast, since the system considered here is 2D, applying the field in-plane (within the $xy$-plane) eliminates orbital effects, so any observed Hall conductance arises solely from the Zeeman effect. 
Despite the field configuration being coplanar, the resulting transverse conductivity retains the antisymmetric tensor structure characteristic of conventional Hall effects. However, unlike the ordinary Hall effect, this planar Hall response—termed the anomalous planar Hall effect—is independent of the relative orientation between the electric and magnetic fields within the plane \cite{PhysRevResearch.3.L012006,PhysRevB.102.241105}. Furthermore, since symmetry ensures that the Berry curvature monopole (and consequently the anomalous planar Hall effect) vanishes at zero magnetic field, the presence of the anomalous planar Hall effect can be experimentally discerned by comparing responses in the presence and absence of an applied field.

We also note that, in principle, an additional Berry curvature-induced contribution to the planar Hall effect, proportional to $\bm{v_\bm{k}} \cdot \bm{\Omega_\bm{k}}$, can arise when electric and magnetic fields are coplanar, as discussed in prior works \cite{PhysRevLett.119.176804,PhysRevB.100.195113}. However, in two-dimensional systems such as the one studied here, the electron group velocity $\bm{v_\bm{k}}$ is always orthogonal to the Berry curvature $\bm{\Omega_\bm{k}}$, rendering this term identically zero. Therefore, the planar Hall response in our system is solely governed by the Berry curvature integral in Eq.~(\ref{eq:AHE}), reflecting the quantum topological nature of the response.
Now, the anomalous planar thermal Hall effect refers to the Berry curvature-induced in-plane transverse thermal response to applied coplanar magnetic field $\bm{B}$ and thermal gradient ($-\bm{\nabla} T$). The corresponding conductivity, designated as $\kappa_{xy}$ in Eq.~(\ref{eq:currents}), is given at a finite temperature $T$ by \cite{PhysRevLett.104.066601,PhysRevB.83.161407} 
\begin{equation}
\begin{split}
     \kappa_{xy} =\frac{k^2_{B}T}{\hbar}\int _{\bm{k}}& \Omega_{\bm{k}}^{n}\biggl[\frac{(\epsilon_{\bm{k}}^{n}
    -\mu)^2}{(k_{B}T)^2} f^{n}_{\bm{k}} + \frac{\pi^2}{3} + \dots \\&- \ln^2(1+e^{-\beta(\epsilon_{\bm{k}}^{n}-\mu)}) -2\textrm{Li}_2(1-f^{n}_{\bm{k}}) \biggr],
\end{split}
\end{equation}
where $\textrm{Li}_2(z)$ is the dilogarithm function and $\beta = k_BT$ with $k_B$ being the familiar Boltzmann constant. Integrating by parts, it is possible to recast the above expression in terms of the zero temperature anomalous planar Hall conductivity as \cite{PhysRevLett.104.066601,PhysRevLett.107.236601,PhysRevB.83.161407}:
\begin{equation}\label{eq:thermalHall_E}
    \kappa_{xy} = -\frac{1}{e^2 T}\int_{-\infty}^{\infty} d\epsilon (\epsilon-\mu)^2 \sigma^{0}_{xy}(\epsilon)f'(\epsilon),
\end{equation}
where $\sigma^{0}_{xy}(\epsilon)$ is the zero temperature anomalous planar Hall conductivity. Here we use $\int \delta(\epsilon-\epsilon_{\bm{k}}^{n})d \epsilon = 1$, $f'(\epsilon-\epsilon_{\bm{k}}^{n})|_{T=0} = -\delta(\epsilon-\epsilon_{\bm{k}}^{n})$,  
$\frac{\partial}{\partial \epsilon} \textrm{Li}_{2}(1-f^{n}_{\bm{k}})=-\beta f^{n}_{\bm{k}} \ln (f^{n}_{\bm{k}})$, and  $\textrm{Li}_{2}(1) =  \pi^2/6$.

The anomalous planar Nernst effect refers to the generation of a transverse charge current in the presence of applied coplanar magnetic field $\bm{B}$ and thermal gradient $\bm{-\nabla T}$ due to a Berry-phase-induced intrinsic mechanism. The corresponding conductivity $\alpha_{xy}$ for a finite temperature $T$ is given by~\cite{PhysRevLett.104.066601,PhysRevB.83.161407,PhysRevLett.97.026603} 
\begin{equation}
\begin{split}
     \alpha_{xy} = -\frac{1}{T}\frac{e}{\hbar} \int_{\bm{k}}& \Omega_{\bm{k}}^{n}\biggl[(\epsilon^n_{\bm{k}}-\mu)f + \dots\\ &+ k_{B}T\ln{(1+e^{-\beta(\epsilon^n_{\bm{k}}-\mu)})}\biggr].   
\end{split}
\end{equation}
Similarly, the anomalous planar Nernst conductivity can be obtained from the zero temperature anomalous planar Hall conductivity ~\cite{PhysRevLett.104.066601,PhysRevB.83.161407,PhysRevLett.97.026603} as:
\begin{equation}\label{eq:nernst_E}
    \alpha_{xy} = -\frac{1}{eT}\int d\epsilon (\epsilon-\mu) \sigma^{0}_{xy}(\epsilon)f'(\epsilon).
\end{equation}

The finite-temperature anomalous planar Nernst conductivity and thermal Hall conductivity are directly related to the zero-temperature anomalous planar Hall conductivity $\sigma^0_{xy}(\epsilon)$ via the Mott relation and the Wiedemann-Franz law respectively, which can be obtained by Sommerfeld expansion~\cite{Ashcroft_1976,PhysRevLett.97.026603} as:
\begin{equation}\label{eq:WFM}
    \alpha_{xy} = eL_0T\frac{\partial\sigma_{xy}}{\partial\mu}, \quad \kappa_{xy} = L_0T\sigma_{xy}.
\end{equation}
Here, $L_0 = \pi^2k_{B}^2/3e^2$ is the Lorentz number. The Mott relation reveals that the anomalous planar Nernst conductivity is a Fermi surface quantity, whereas the Wiedemann–Franz law implies that the anomalous planar thermal Hall conductivity is determined by contributions from the entire Fermi sea.
 
\section{\label{sec: model}Model Hamiltonian}
We consider a two-dimensional single orbital model characterizing a $d$-wave altermagnet on a square lattice with lattice constant a = 1, which can be written as ~\cite{Yan_2023}:
\begin{eqnarray}\label{eq:hamTM}
    H^{\rm AM} = &-2t(\cos k_x + \cos k_y) + 2\lambda(\sin k_y \sigma_x - \sin k_x \sigma_y) \nonumber \\
    &+ 2t_{\text{AM}}(\cos k_x - \cos k_y)\sigma_z,
\end{eqnarray}
where $t$ is the nearest-neighbor hopping parameter, t$_{\rm AM}$ denotes the d$_{x^2-y^2}$ altermagnetic order parameter and  $\lambda$ represents the strength of the Rashba spin-orbit coupling (RSOC). $\sigma_{0}$ is the $2\times2$ identity matrix and the Pauli matrices $\bm{\sigma}=(\sigma_x,\sigma_{y},\sigma_{z})$ act in spin space. The band structure of the model system in Eq.~(\ref{eq:hamTM}) is depicted in Fig.~\ref{fig:hamiltonian}(a) showing a gapless Dirac cone at the $\Gamma$ point. \textcolor{black}{In the context of symmetries, the Hamiltonian belongs to the magnetic point group $4'm'm$, which is generated by $\hat{\mathcal{C}}_{4z}\hat{\cal{T}}$and $\hat{\mathcal{M}}_x\hat{\cal{T}}$. While time-reversal symmetry (TRS; ${\hat{\cal T}} = -i\sigma_y {\hat{\cal K}}$ where $\hat{\cal{K}}$ is the complex conjugate), four-fold rotation symmetry about the $z$-axis ($\hat{\mathcal{C}}_{4z}=e^{i\frac{\pi}{4}\hat{\sigma}_z}$), and reflection symmetry across the $x$-axis ($\hat{\mathcal{M}}_x=i\hat{\sigma}_{x}$) are individually broken, the Hamiltonian remains invariant under the combined operators $\hat{\mathcal{C}}_{4z}\hat{\cal{T}}$and $\hat{\mathcal{M}}_x\hat{\cal{T}}$. These generators imply the existence of two additional mirror symmetries: $\hat{\mathcal{M}}_{x=y}=(\hat{\mathcal{M}}_x\hat{\cal{T}})(\hat{\mathcal{C}}_{4z}\hat{\cal{T}})$ and $\hat{\mathcal{M}}_{x=-y}=(\hat{\mathcal{C}}_{4z}\hat{\cal{T}})(\hat{\mathcal{M}}_x\hat{\cal{T}}).$ Both the $\hat{\mathcal{C}}_{4z}\hat{\cal{T}}$ symmetry and the mirror symmetries mandate a vanishing Berry curvature monopole. Specifically, $\hat{\mathcal{C}}_{4z}\hat{\mathcal{T}}$ enforces the transformation $\Omega(k_x,k_y)\rightarrow - \Omega(-k_y,k_x)$, while $\hat{\mathcal{M}}_{x=y}$ requires $\Omega(k_x,k_y)\rightarrow - \Omega(-k_y,-k_x)$  and thereby either symmetry prohibits a linear anomalous Hall response. For a finite monopole, $\hat{\mathcal{C}}_{4z}\hat{\mathcal{T}}$ and both mirror symmetries need to be broken simultaneously. In this work, we accomplish that symmetry breaking by introducing an external magnetic field as discussed below.} The anomalous thermal Hall and anomalous Nernst effects follow the same symmetry constraints as the AHE and are consequently expected to be finite under the same conditions.

In the presence of a Zeeman field with arbitrary orientation, the altermagnetic system described by the Eq.~(\ref{eq:hamTM}) can be written as:
\begin{equation}\label{eq:hamLat}
    H(\bm{k}) = H^{\rm AM}(\bm{k}) - \bm{b}\cdot\bm{\sigma},
\end{equation}
where $\bm{b} = \frac{g\mu_B}{2}\bm{B}$, with $\mu_{B}$ and $g$ being the Bohr magneton and the effective $g$-factor (assumed to be isotropic), respectively.    \textcolor{black}{The Zeeman term $\bm{b\cdot\sigma}$ always breaks $\hat{C}_{4z}\hat{\mathcal{T}}$ and can induce a finite monopole provided the mirror symmetries are also broken.}
To investigate the anomalous planar responses (Hall, thermal Hall, and Nernst effects), we first consider the system in Eq.~(\ref{eq:hamLat}) subject to an in-plane magnetic field $\bm{B_{\parallel}} = (B_x,B_y) = (B_{\parallel}\cos\phi,B_{\parallel}\sin\phi)$ where $\phi$ is the azimuthal angle. \textcolor{black}{We note that when $\bm{B}$ is aligned along $k_x = \pm k_y$ (equivalently when $B_x=\pm B_y$), the in-plane Zeeman term $b_x\sigma_x+b_y\sigma_y$ is invariant under $\hat{\mathcal{M}}_{x=\pm y}$ and consequently, the planar responses vanish.} We then extend our discussion to include an out-of-plane component ${B_{\perp}} = B_z = B\cos\theta$. \textcolor{black}{In the present analysis, we assume that the system remains in its altermagnetic ground state and does not undergo a spin-flop-like transition driven by the Zeeman field, since the applied in-plane field ($b = 0.002$ eV) is much smaller than the altermagnetic order parameter ( $t_{\text{AM}}$ = 0.25 eV).}
\begin{figure}[!t]
\centering
\includegraphics[width=0.48\textwidth]{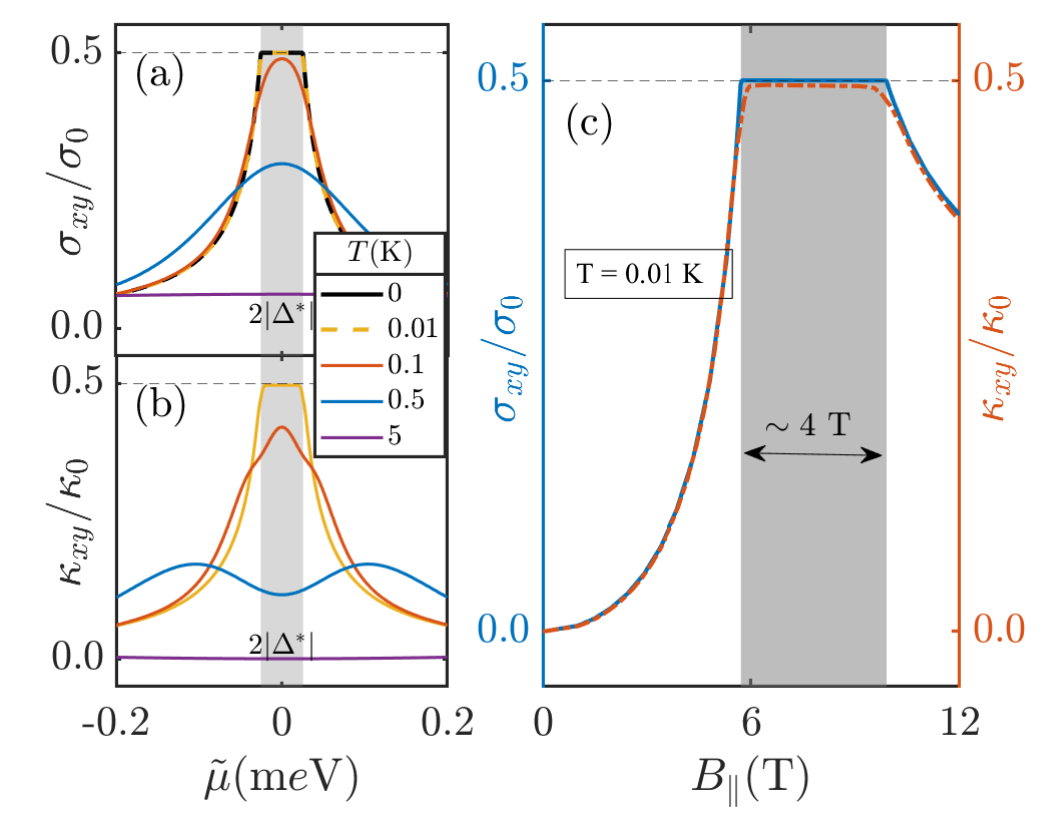}
\caption{Numerically calculated amplitudes for intrinsic anomalous planar charge and thermal Hall responses for the model Hamiltonian (\ref{eq:hamLat}) using the same parameters as those mentioned in the caption of Fig.~\ref{fig:hamiltonian}. Panel (a) and (b) show the magnitude of anomalous planar Hall conductance $\sigma_{xy}$ and thermal Hall conductance $\kappa_{xy}$ as a function of shifted chemical potential $\tilde{\mu}=\mu-\epsilon^*_{7\text{T}}$ with $\epsilon^* = -t(4-\bm{k^{*2}})$, respectively, at various temperatures for a planar magnetic field $B_y = 7$ T. (c) The magnitude of both $\sigma_{xy}$ and $\kappa_{xy}$ at the shifted chemical potential $\tilde{\mu} = 0$ are plotted as a function of applied planar magnetic field $B_{\parallel}$ at a fixed in-plane angle $\phi = \pi/2$ with $T = 0.01$ K, showing approximately half-integer quantized anomalous planar thermal Hall plateau in $d$-wave altermagnets. Here, $\sigma_0 = e^2/h$ and $\kappa_0 = (\pi^2/3)k_B^2T/h$.}
\label{fig:SigmaKappa}
\end{figure}

\begin{figure*}[!htb]
\centering
\includegraphics[width=0.98\textwidth]{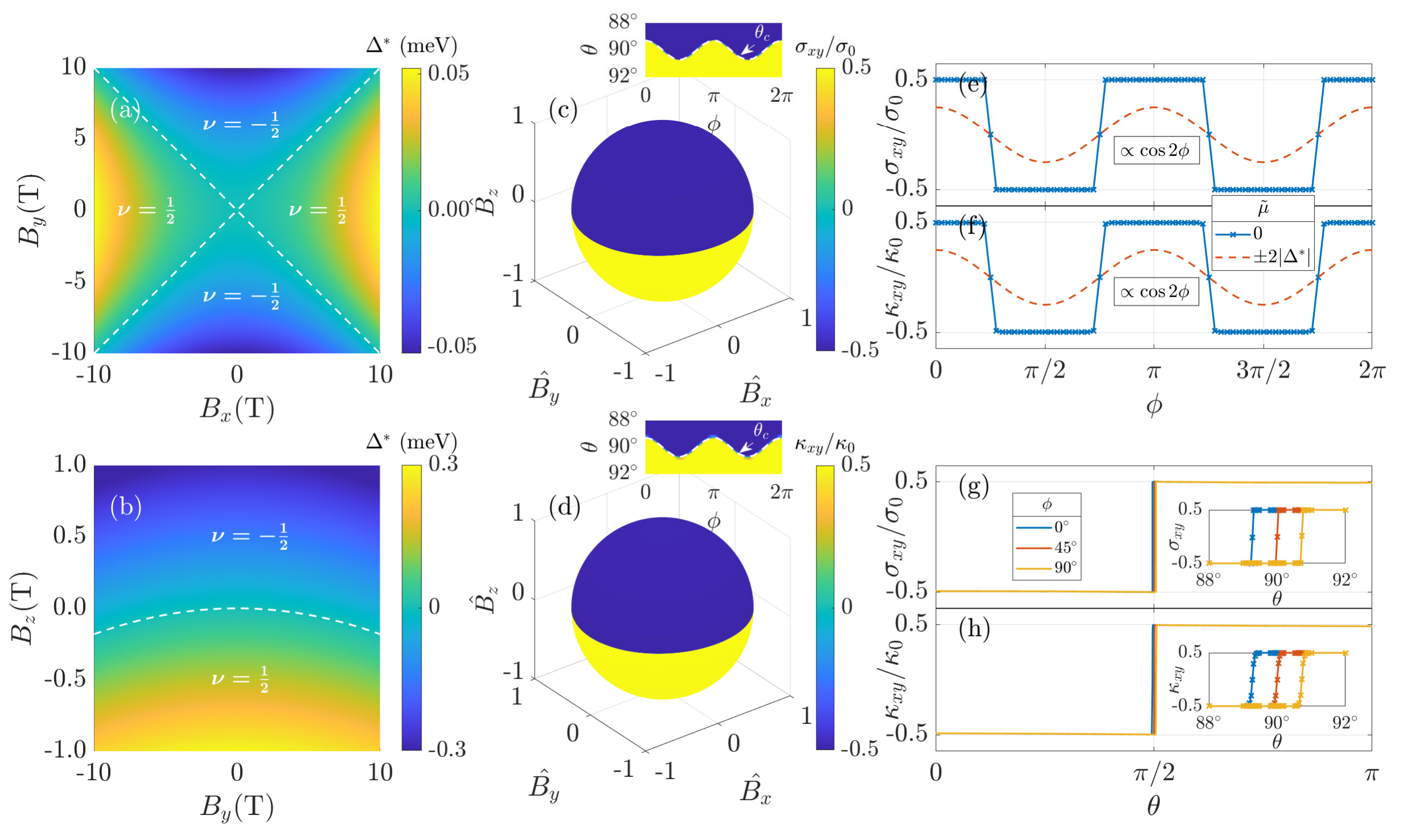}
\caption{Phase diagram for the $\mathcal{C}_{4z}\mathcal{T}$ breaking $d$-wave altermagnet model given in Eq.~(\ref{eq:hamLat}). Panel (a) shows the Dirac mass term $\Delta^{*}$ (see Eq.~(\ref{eq:gap})) as a function of $B_x$ and $B_y$ for a purely in-plane magnetic field and describes the dependence of the topological phase on the in-plane angle $\phi=\arctan(B_y/B_x)$. The period of $\pi$ emerges from the $\cos(2\phi)$ dependence of $\Delta^{*}$. The white dashed lines show the band gap closing as a result of $\hat{\mathcal{M}}_{x=\pm y}$ symmetry being restored and indicate a phase transition. The sign of the \textcolor{black}{BCM} $\nu$ is specified for each quadrant, revealing a period of $\pi$. In panel (b), $\Delta^*$ is shown as a function of in-plane component $B_y$ and out-of-plane component $B_z$. The dashed white curve shows the band gap closing and separates the two phases with different signs for the conductivities. In panels (c) and (d), the distinct signs of the two topological phases becomes manifest via the signs of the nearly half-quantized anomalous planar charge Hall and thermal Hall responses respectively at $T = 0.01$ K. The insets of panels (c) and (d) show that the topological phase boundary at any given $\phi$ does not lie exactly on the equator but is rather dictated by the critical angle $\theta_c$ given by Eq.~(\ref{eq:critang}). 
The planar anomalous charge conductivity $\sigma_{xy}$ and thermal Hall conductivity $\kappa_{xy}$ depicted in panels (e) and (f), respectively where the magnetic field is restricted to the $xy$-plane but the azimuthal angle is varied at $T=0.01$ K. Panels (g) and (h) show the variation of $\sigma_{xy}$ and $\kappa_{xy}$ as an out-of-plane component of the magnetic field is introduced by varying the polar angle $\theta$. The parameters used are the same as those mentioned in the caption of Fig.~\ref{fig:hamiltonian}. Here, $\sigma_0 = e^2/h$ and $\kappa_0 = (\pi^2/3)k_B^2T/h$}
\label{fig:angular_dep}
\end{figure*}

To analyze the band structure evolution under an in-plane magnetic field (depicted in Fig.~\ref{fig:hamiltonian}(d)), we expand Eq.~(\ref{eq:hamLat}) near the $\Gamma$ point:
\begin{equation}
H(\bm{k}) = t({k}^2-4)+\bm{d(k) \cdot\sigma},
\end{equation}
with $\bm{d(k)}=[2\lambda k_y - b_x,-2\lambda k_x - b_y,t_{\text{AM}}(k_y^2-k_x^2)-b_z]$. An in-plane magnetic field shifts the massless Dirac cone at the $\Gamma$-point to a tilted Dirac cone at $\bm{k}^*\equiv \bm{\hat{z}}\times\frac{\bm{b_{\parallel}}}{2\lambda}$ where $\bm{b_{\parallel}}=(b_x, b_y)$. This shift introduces a finite mass through the altermagnetic term by allowing $d_{z}(\bm{k})$ to be non-zero at the Dirac point even in the absence of $\bm{b}_{\perp}\equiv b_z$. Here, the energy spectra of the two bands are
\begin{equation}
    \epsilon_{\pm} =  -t(4 -{\bm{k}}^2) \pm |\bm{d(k)|},
\end{equation} 
where $\pm$ refer to the upper and lower bands respectively.
At the shifted Dirac point $\bm{k}^*$, the Zeeman-induced gap $2|\Delta^*|$ is determined by the mass term: 
\begin{equation}\label{eq:gap}
\Delta^* = t_{\text{AM}}\frac{b^2_{\parallel}}{(2\lambda)^2}\cos2\phi -b_z.   
\end{equation} 
The closing of the Dirac gap signifies a topological phase transition (see Fig.~\ref{fig:angular_dep}(a) and (b)). When a small out-of-plane component is present ($B_z<<B$), this transition occurs at a critical angle $\theta_c$ close to $\pi/2$ for a given $\phi$: 
\begin{equation}\label{eq:critang}
\theta_c=\arccos\biggl(\frac{t_{\text{AM}}}{(2\lambda)^2}b\cos{2\phi}\biggr).
\end{equation}
Expanding the Hamiltonian near 
$\bm{k}^*$ to linear order in $\tilde{\bm{k}}$, we obtain 
\begin{equation}\label{eq:LinHam}
\begin{split}
    H^*(\bm{k}) \approx & \epsilon^*\sigma_{0} + 2t\bm{k^*\cdot\tilde{k}}\sigma_{0} + 2\lambda\tilde{k_y}\sigma_{x}-2\lambda\tilde{k_x}\sigma_{y}\\   &\Delta^*\sigma_z + 2t_{\text{AM}}(k_y^*\tilde{k_y}-k^*_{x}\tilde{k_x})\sigma_{z},
\end{split}
\end{equation}
where $\bm{\tilde{k}} = \bm{k}-\bm{k}^*$ is the in-plane wave vector measured from the shifted Dirac point, and the midpoint of the Dirac energy gap is given by:
\begin{equation}\label{eq:midgap}
    \epsilon^* = -t(4-\bm{k}^{*2}).
\end{equation}
\textcolor{black}{In addition to the Dirac cone at the $\Gamma$-point, the Hamiltonian Eq.~(\ref{eq:hamTM}) hosts a Dirac cone at $\mathrm{M} = (\pi,\pi)$. In the presence of an in-plane magnetic field, a similar analysis shows that the Dirac point is shifted to $(\pi-k^*_x,\pi-k^*_y)$, midpoint of the Dirac energy gap is given by $t(4-\bm{k}^{*2})$, and the magnitude of the gap remains the same. In this work, we focus on the regime $t>>|\Delta^*|$, where the energy separation between the Dirac cones at $\Gamma$ and at $\mathrm{M}$ is significantly larger than the Dirac gaps. In this case, tuning the chemical potential to the midpoint of the massive Dirac cones near $\Gamma$ ensures that the Berry curvature contribution from $
\mathrm{M}$ is absent, yielding approximate half-quantization, as we elaborate in Sec. \ref{sec: results}.}

In 2D, the Berry curvature distribution for a two-band Hamiltonian can be calculated by the following formula \cite{PhysRevB.74.085308}:
\begin{equation}
    \Omega_{\pm}(\bm{k}) = \mp\frac{1}{2}\hat{\bm{d}}\cdot(\partial_{k_x} \hat{\bm{d}} \times \partial_{k_y} \hat{\bm{d}}).
\end{equation}
Here, $\bm{\hat{d}(k)}$ specifies the spin texture. Using this expression, the Berry curvature for the linearized Hamiltonian in Eq.~(\ref{eq:LinHam}), localized near the anticrossing point $\bm{k}^*$, can be obtained to leading order in $\tilde{\bm{k}}$ as:
\begin{equation}
    \Omega^{*}_{\pm}(\bm{k})\approx \mp\frac{2\lambda^2\Delta^*}{\bigl[4\lambda^2\tilde{k}^2+(\Delta^* + 2t_{\text{AM}}(\tilde{k_y}k_y^{*}-\tilde{k_x}k_x^{*}))^{2}\bigr]^{3/2}}.
\end{equation}
 The Berry curvature distribution of the conduction band corresponding to the Hamiltonian in Eq.(\ref{eq:hamLat}) is shown in Fig.~\ref{fig:hamiltonian}~(c) and (f), in the absence and presence of an in-plane field, respectively. In the absence of a magnetic field, the spin texture seen in Fig.~\ref{fig:hamiltonian}~(b) follows the $\hat{\mathcal{C}}_{4z}\hat{\mathcal{T}}$ symmetry as expected, and the Berry curvature distribution has a four-lobed quadrupolar structure centered at the $\Gamma$ point which hosts the Dirac node. With equal contributions from the positive and negative lobes of the distribution, the BCM vanishes, resulting in a vanishing linear anomalous Hall response while allowing for a finite third-order Hall effect as a leading-order~\cite{Fang_2024,Zhang_2023,PhysRevB.110.195119}. In contrast, when $\bm{b}_{\parallel}$ is applied, the spin texture $\bm{\hat{d}}$ reflects a broken $\hat{C}_{4z}\hat{\mathcal{T}}$ symmetry, as evidenced by the unequal distribution of red and blue regions that represent spin polarization in the plane, as seen in Fig.~\ref{fig:hamiltonian}(e). Additionally, the out-of-plane spin component, indicated by the arrows, varies in magnitude along the Fermi surface and is not consistently tangential, reflecting a broken $\hat{\mathcal{M}}_{x=\pm y}$ symmetry. This modification of spin textures, together with the gap opening at the shifted Dirac point $\bm{k}^*$, leads to a well-defined Berry curvature throughout the Brillouin zone. As shown in Fig.~\ref{fig:hamiltonian}(f), this gives rise to a monopole-like distribution centered at the massive Dirac node, resulting in a finite linear anomalous planar Hall conductivity, as we will elaborate in the next section.

\section{\label{sec: results}Results}
With the Berry curvature in hand, we proceed to evaluate the anomalous planar transport coefficients at a realistic in-plane magnetic field strength applied along the $y$-direction, $\bm{B}_{\parallel} \equiv B_y = 7$ T. Throughout this section, we employ a shifted chemical potential, 
\begin{equation}\label{eq:scp}
\tilde{\mu} = \mu - \epsilon^*_{7\text{T}}, 
\end{equation}
defined with respect to $\epsilon^*_{7\text{T}}$ which is the midpoint of the Dirac gap (see Eq.~(\ref{eq:midgap}))  at $B_{x} = 7$ T. Figs.~\ref{fig:SigmaKappa}(a) and (b) display the anomalous planar Hall and thermal Hall conductivities as a function of  $\tilde{\mu}$ at various temperatures. It is evident from Fig.~\ref{fig:SigmaKappa}(a) that at zero temperature, when the chemical potential lies within the gap $2|\Delta^*|$ opened at the shifted Dirac point $\bm{k^{*}}$, the anomalous planar Hall conductivity (APHC) becomes half-quantized. Specifically, it is approximated to $\sigma_{xy} \approx \text{sgn}(\Delta^*) \frac{e^2}{2h}$. It is attributed to the fact that in this regime, the Berry curvature is well-defined and peaks at the Dirac anti-crossing point, and the Fermi surface encloses a large loop in the lower band, well separated from $\bm{k}^*$. With the outer Fermi surface $k_F\gg|\Delta^{*}|/\lambda$, the integration of the Berry curvature over occupied states evaluates to $\pi$ as discussed in \cite{PhysRevB.68.045327}. This is precisely the means through which half-quantization is achieved. As $\tilde{\mu}$ is tuned below the gap, the states near $\bm{k}^*$ in the lower band—which contribute most significantly to the APHC—become depopulated, thereby decreasing the Hall response. In contrast, when $\tilde{\mu}$ moves above the gap, the states near $\bm{k}^*$ in the upper band become occupied. The Berry curvature of this band, having an equal magnitude but opposite sign, offsets the contribution from the lower band, leading to suppression of the net APHC. The anomalous planar thermal Hall conductivity (APTHC)  exhibits behavior similar to APHC as seen in Fig.~\ref{fig:SigmaKappa}(b). It should be noted that both the electrical conductivity $\sigma_{xy}$ and the thermal conductivity $\kappa_{xy}$ remain even functions of the shifted chemical potential $\tilde{\mu}$. 
\begin{figure}[!t]
\centering
\includegraphics[width=0.48\textwidth]{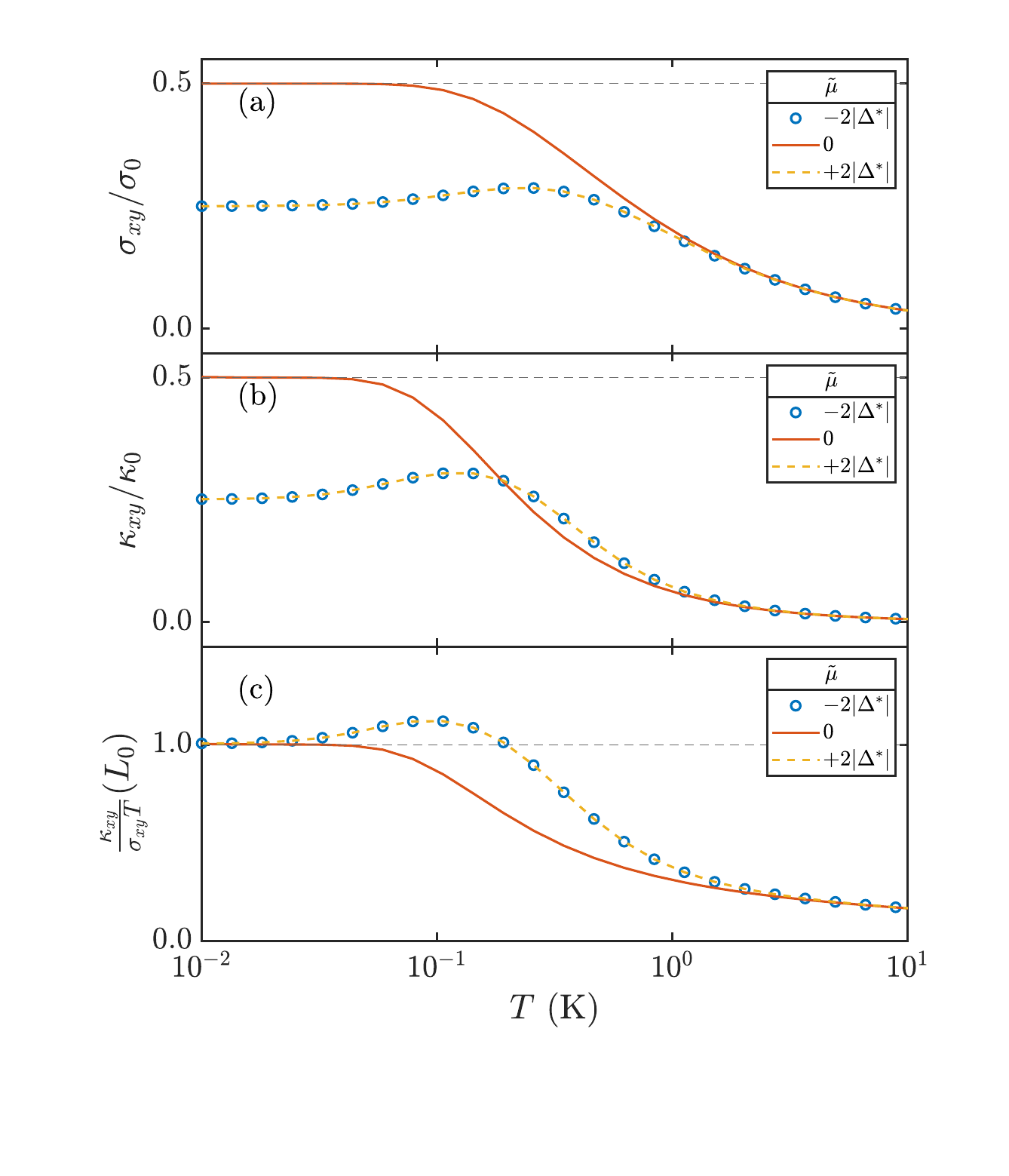}
\caption{The variation of $\sigma_{xy}$ and $\kappa_{xy}$ with temperature at three different chemical potentials for planar magnetic field $B_x = 7$ T is depicted in panels (a) and (b) respectively. The chemical potentials considered here correspond to the center of the Dirac band gap and to points symmetrically located in the upper and lower bands, represented by the shifted chemical potentials $\tilde{\mu}=0,\pm2|\Delta^*|$. Panel (c) shows the validity of the Wiedemann Franz law at low temperatures for all the chemical potentials and a rapid departure from it at temperatures higher than $0.1$ K, i.e. the temperature corresponding to the magnitude of the Dirac band gap energy. All the other parameters are same as Fig.~\ref{fig:hamiltonian}. Here, $\sigma_0 = e^2/h$ and $\kappa_0 = (\pi^2/3)k_B^2T/h$.}
\label{fig:tempDep}
\end{figure}
The dependence of APHC and APTHC on the strength of the in-plane magnetic field at $\tilde{\mu}=0$ (or equivalently, $\mu = \epsilon^{*}_{7\text{T}}$) is shown in Fig.~\ref{fig:SigmaKappa}(c). We find that both APHC and APTHC exhibit half-quantized plateaus over a range of magnetic fields. It is because the in-plane field $B_\parallel$ modifies the shifted Dirac node momentum $\bm{k}^*$ ($=\frac{\bm{\hat{z}\times b_{\parallel}}}{2\lambda}$), which in turn increases both the width of the gap $2|\Delta^{*}|$ (see Eq.~(\ref{eq:gap})) and the energy $\epsilon^*$ (see Eq.~(\ref{eq:midgap})) where the gap is centered. The half-quantized plateaus thus appear over a range of magnetic fields (approximately $6-10$~T at $T = 0.01$ K) as shown in Fig.~\ref{fig:SigmaKappa}(c) for which $\epsilon^{*}_{7\text{T}}$ lies within the gap, provided $T<<|\Delta^{*}|/k_B$. When the magnetic field exceeds 10 T, $\epsilon^{*}_{7\text{T}}$ moves outside the gap, and the quantization breaks down due to contributions from the upper band.        
\begin{figure*}[!ht]
\centering
\includegraphics[width=0.99\textwidth]{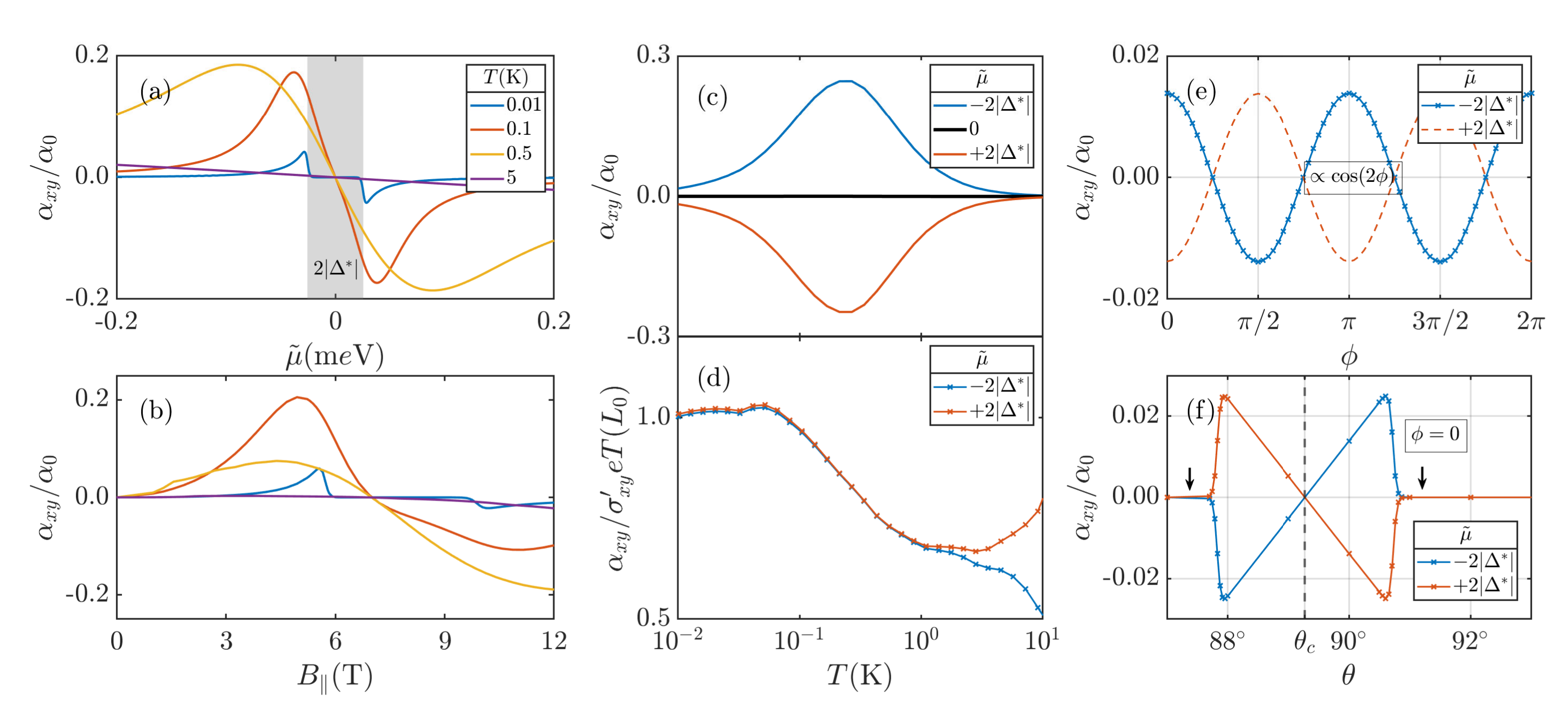}
\caption{Numerically calculated anomalous planar Nernst conductivity for the model Hamiltonian (\ref{eq:hamLat}) using the same parameters as those mentioned in the caption of Fig.~\ref{fig:hamiltonian}. Panel (a) shows the magnitude of anomalous planar Nernst conductivity $\alpha_{xy}$ in units of $\alpha_0 = ek_B/hT$ plotted against the shifted chemical potential $\tilde{\mu}$ (see Eq.~(\ref{eq:scp})) at various temperatures for a planar magnetic field $B_x = 7$ T. Panel shows (b) variation of $\alpha_{xy}$ with the strength of in-plane magnetic field. Panel (c) shows the variation of $\alpha_{xy}$ with temperature at chemical potentials corresponding to the center of the Dirac band gap and to points symmetrically located in the upper and lower bands, represented by the shifted chemical potentials $\tilde{\mu}=0,\pm2|\Delta^*|$. Panel (d) depicts that the Mott relation holds for low temperatures; $\sigma'_{xy}\equiv\partial\sigma_{xy}/\partial\mu$. Panel (e) shows the expected $\cos(2\phi)$ dependence on the azimuthal angle of $\bm{B}_{\parallel}$ at $T = 0.01$ K. Panel (f) demonstrates how quickly (i.e. at small deviations from $\theta = \pi/2\equiv90^\circ$) $\alpha_{xy}$ is made negligible when an out-of-plane component of magnetic field is introduced at $\phi = 0$ and $T = 0.01$ K.}
\label{fig:nernst}
\end{figure*}

To explore this system from a topological perspective, we plot the Dirac mass term $\Delta^*$ at $\bm{k}^*$ as a function of $B_x$ and $B_y$ for a purely in-plane magnetic field $\bm{B_{\parallel}}$ in Fig.~\ref{fig:angular_dep}. Fig \ref{fig:angular_dep}(a) illustrates the $\cos 2\phi$ dependence of the Dirac band gap at $\bm{k}^*$ in the absence of an out-of-plane magnetic field component. The closing of the gap signals a topological phase transition, separating phases characterized by two distinct values of the \textcolor{black}{BCM} $\nu = \pm 1/2$ (corresponding to the lower band) with a periodicity of $\pi$, consistent with the findings of Ref.~\cite{Rao_2024}. With $B_z = 0$, Eq.~(\ref{eq:gap}) indicates that the gap closes at $\cos2\phi =0$ or at $B_x = \pm B_y$ which corresponds to the boundary lines indicated in Fig.~\ref{fig:angular_dep}(a). When an out-of-plane component is included, with $B_x = 0$, the gap closes along the parabola $B_z\propto -B_y^2$ as depicted in Fig.~\ref{fig:angular_dep}(b). It is important to note that in systems where the Berry curvature integrated over the Brillouin zone  yields a nonzero \textcolor{black}{BCM} $\nu$, the anomalous Hall conductivity is quantized as $\sigma_{xy} = \nu \sigma_0$, with $\sigma_0 = e^2/h$ denoting the Hall conductance quantum. In the present case, due to the symmetry-breaking Zeeman gap at the shifted Dirac point, the system exhibits a half-integer \textcolor{black}{BCM}, resulting in a nearly half-quantized APHC.  Through the Wiedemann Franz law in Eq.~(\ref{eq:WFM}), this immediately leads to a nearly quantized $\kappa_{ab}$ at low temperatures,
$\kappa_{xy}=\nu \kappa_{0}$, where $\kappa_0 = (\pi^2/3)k^2_{B}T/h$ as the thermal Hall conductance quantum. The full angular dependence of APHC and APTHC is illustrated in panels (c) and (d) of Fig.~{\ref{fig:angular_dep}}.  Eq.~(\ref{eq:gap}) defines the boundary where $\Delta^{*}$ changes sign, marking the topological phase transitions. This boundary is given by the critical angle $\theta_c$ (see Eq.~(\ref{eq:critang})), which accounts for the observed $\cos(2\phi)$ dependence around the equator in the insets of panels (c) and (d) in Fig.~\ref{fig:angular_dep}. At small magnetic fields ($b<<4\lambda^2/t_{\text{AM}})$), the contribution from $b_{\parallel}$ becomes negligible at polar angles away from the equator since $\Delta^{*}$ depends linearly on $b_z \propto b\cos\theta$ and quadratically on $b_{\parallel} = b\sin\theta$. Consequently, the azimuthal angular dependence fades, and the \textcolor{black}{BCM} is determined solely by the sign of $b_z$. For a clearer interpretation of the angular dependence of both APHC and APTHC, we decompose these responses into their polar and azimuthal components . In Fig.~\ref{fig:angular_dep}(e) and (f), the topological phase transitions and corresponding changes in the \textcolor{black}{BCM} manifest as quantized jumps in the anomalous planar Hall and thermal Hall conductivities at low temperatures. Fig.~\ref{fig:angular_dep}(e) and (f) shows the first order transverse responses vanishing at $\phi = (2n+1)\frac{\pi}{4}$ where the $\hat{\mathcal{M}}_{x=\pm y}$ symmetry is restored and BCM vanishes. In this scenario with $d_z(\bm{k}^*) = 0$, the spin textures lie in the $xy$ plane with the spins oriented perpendicular to the wave vector. This leads to the Berry curvature being singular at the band-touching point and vanishing everywhere else and does not provide a contribution to the APHC. As for the polar angle $\theta$, Fig \ref{fig:angular_dep} (g) and (h) show that reversing the direction of $B_z$ flips the sign of $\sigma_{xy}$ and $\kappa_{xy}$, while the magnitude of the half-quantized conductivities does not vary appreciably apart from the discontinuity at the polar angles near $\pi/2$ dictated by $\phi$ through Eq.~(\ref{eq:critang}). The anomalous planar Hall effects are even in $B_{\parallel}$ and odd in $B_{z}$. We note that the relative angle between the driving electric field $\bm{E}$ (or the applied thermal gradient $-\nabla{T}$) and the applied magnetic field does not enter the expression for Berry curvature and therefore does not affect the response. Specifically, while the  anomalous planar  transport coefficients depend on the orientation of $\bm{B_{\parallel}}$, they are insensitive to the relative angle between $\bm{B_{\parallel}}$ and $\bm{E_{\parallel}}$ ($-\nabla_{\parallel}T$).

Finally, the temperature dependence of the APHC and APTHC is depicted in Fig.~\ref{fig:tempDep}. In Fig.~\ref{fig:tempDep} (a) and (b), we find that the characteristic half-quantization persists only at temperatures well below $T\sim0.35$~K. This limitation arises due to thermal broadening of the Fermi-Dirac distribution, which induces contributions from the upper band even when the chemical potential $\mu$ lies within the gap. As the temperature increases, the range of 
$\mu$ values within the gap that support the half-quantized response becomes narrower, resulting in a shrinking plateau width and a suppression of the maximum amplitude—consistent with the trends observed in Fig.~\ref{fig:SigmaKappa}(a)–(b). With a comparison of the amplitudes and profiles of $\sigma_{xy}$ and $\kappa_{xy}$ in Fig \ref{fig:SigmaKappa} (a) and (b), the rapid departure from the Wiedemann-Franz law with increasing temperature at any chemical potential becomes evident, as illustrated in \ref{fig:tempDep}(c). The Lorenz number $L = \kappa_{xy}/\sigma_{xy}T$ is generally less than $L_0$ at higher temperatures as expected. The overlap in the behavior of  $\kappa_{xy}/\sigma_{xy}T$ at $\tilde{\mu} = \pm2|\Delta^*|$ arises from the fact that both $\sigma_{xy}$ and $\kappa_{xy}$ exhibit symmetric profiles around $\tilde{\mu}=0$ as seen in Fig.~\ref{fig:SigmaKappa}~(a) and (c).

In Fig.~\ref{fig:nernst}, we examine the anomalous planar Nernst conductivity $\alpha_{xy}$ for the altermagnetic system Eq.~(\ref{eq:hamLat}). Apart from Fig.~\ref{fig:nernst}(b) where dependence on magnetic field strength is discussed, the in-plane field is fixed at $B = 7$ T throughout,  keeping the midpoint of the Dirac band gap (see Eq.~(\ref{eq:gap})) at $\epsilon^*_{7\text{T}}$. In Fig.~\ref{fig:nernst}(a), we start by examining the profile of the $\alpha_{xy}$ as it is varied with the shifted chemical potential $\tilde{\mu}$ (see Eq.~(\ref{eq:scp})). In contrast to the anomalous planar Hall and thermal Hall conductivities where the peaks fall within the Dirac gap (see Fig.~\ref{fig:SigmaKappa}(a) and (b)), $\alpha_{xy}$ is seen to vanish within the gap and have oppositely signed peaks on either side of it. This can be understood in one of two ways. The first is through the Mott relation (see Eq.~(\ref{eq:WFM})) which relates $\alpha_{xy}$ to the derivative of $\sigma_{xy}$ with respect to $\mu$. A quick look at Fig.~\ref{fig:SigmaKappa}(a) reveals that at low temperatures, the derivative of $\sigma_{xy}$ increases with $\mu$ before falling to zero within the gap and then decreasing again, similar to the behavior of $\alpha_{xy}$ with $\tilde{\mu}$ seen here. An alternate way of interpreting this behavior is by recalling that the Nernst conductivity is a Fermi surface property at low temperatures as discussed in Sec.~\ref{sec: formalism}. Consequently, $\alpha_{xy}$ captures the opposite signs of the Berry curvature peaks in the upper and lower bands as the chemical potential sweeps across the vicinity of the Dirac band gap as seen in Fig.~\ref{fig:nernst}(a). At the middle of the gap where the sole Fermi surface in the lower band is far from $\bm{k}^*$, $\alpha_{xy}$ does not effectively sample any Berry curvature and thereby vanishes. Now, with increase in temperature, the peaks are seen to shift away from $\tilde{\mu}=0$. This is because the thermal broadening of the Fermi distribution incorporates contributions from more states, causing the peak to reach its maximum at a lower (higher) chemical potential in the lower (upper) band before decreasing as the thermal window further broadens and includes contributions from the opposite band. Fig.~\ref{fig:nernst}(a) can be understood as each $\tilde{\mu}$ being associated with a characteristic temperature at which the Nernst coefficient peaks. 
\begin{figure}[!t]
\centering
\includegraphics[width=0.48\textwidth]{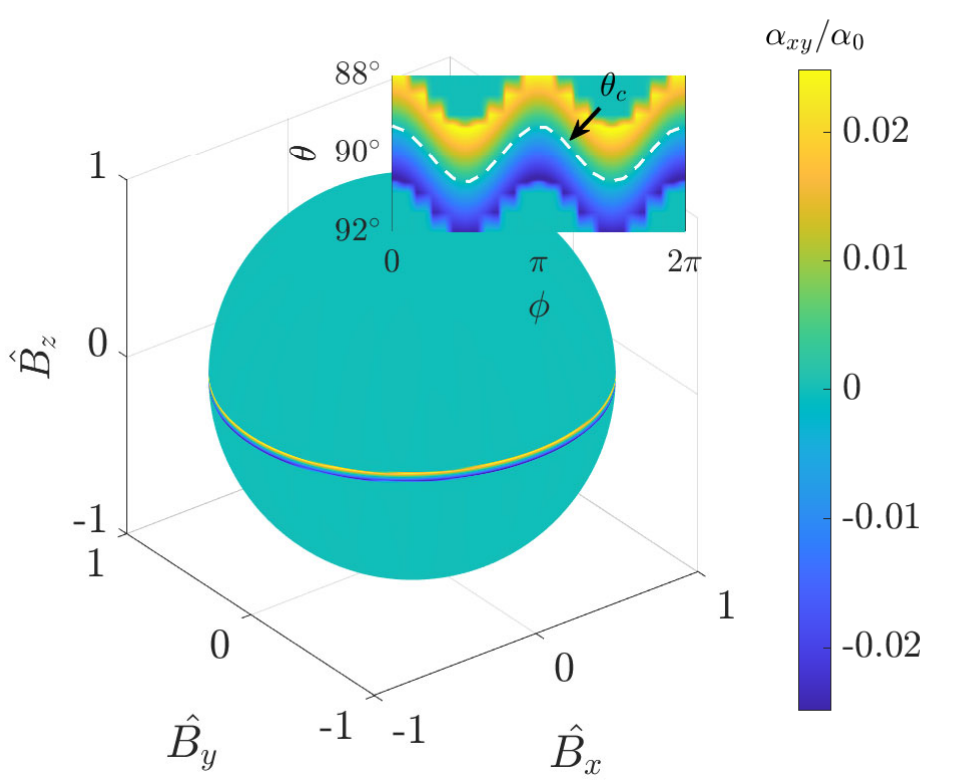}
\caption{The full angular variation of the anomalous Nernst conductivity  $\alpha_{xy}$ for the model Hamiltonian in Eq.~(\ref{eq:hamTM})
using the same parameters as those mentioned in the caption of Fig. \ref{fig:hamiltonian} with a magnetic field strength of
$B = 7$ T at $0.01$ K. 
The planar Nernst conductivity vanishes for most values of $B_z$ except near the equator as the Dirac gap $2|\Delta^*|$ rapidly increases with $B_z$ and the chemical potential falls within the Dirac gap. Since chemical potential within the Dirac gap implies a nearly constant planar anomalous Hall effect, as shown in Fig.~\ref{fig:SigmaKappa} (c) and (d), the planar Nernst vanishes via the Mott relation (Eq.~(\ref{eq:WFM})). The significant variation of $\alpha_{xy}$ happens only near the equator (see text and Fig.~\ref{fig:nernst} (e) and (f). The critical polar angle $\theta_c$ (see Fig.~\ref{fig:nernst}(f)) near the equator is depicted as a function of the azimuthal angle $\phi$ in the inset.}
\label{fig:nernst3D}
\end{figure}
Conversely, in Fig.~\ref{fig:nernst}(c), where $\alpha_{xy}$ is plotted as a function of temperature, we observe a peak at the characteristic temperature corresponding to the chemical potential $\tilde{\mu} = \pm2|\Delta^*|$. The odd-in-$\tilde{\mu}$ profile is reflected in Fig.~\ref{fig:nernst}(c) where the contributions from the lower and upper bands mirror each other with opposite signs. The Mott relation is seen to be followed for temperatures for which $k_BT<< |\Delta^*|$ in Fig.~\ref{fig:nernst}(d). The variation of $\alpha_{xy}$ with the strength of in-plane magnetic field is depicted in Fig.~\ref{fig:nernst}(b). As the in-plane field strength is varied, the magnitude of the Dirac gap $2|\Delta^*|$ (see Eq.(~\ref{eq:gap})) as well as the midpoint of the Dirac gap $\epsilon^*$ (see Eq.~(\ref{eq:midgap})) vary as $B_{\parallel}^2$. Depending on whether the chemical potential, held at a constant $\mu = \epsilon^*_{7\text{T}}$, falls below the gap in the lower band or above the gap in the upper band for a given magnetic field strength, $\alpha_{xy}$ will pick up the corresponding positive or negative peak of the Berry curvature, thus explaining the change in sign of $\alpha_{xy}$. At $B_{\parallel} = 7$ T, the chemical potential is guaranteed to fall within the gap and $\alpha_{xy}$ vanishes for all temperatures (where the temperatures are indicated by the same colors as the ones shown in the legend for Fig.~\ref{fig:nernst}(a)). The in-plane angular dependence of $\cos(2\phi)$ is seen in Fig.~\ref{fig:nernst}(e). Similar to the anomalous Hall and thermal Hall conductivities, the anomalous Nernst conductivity also carries the $\cos(2\phi)$ signature of the system and vanishes when the Dirac gap closes. When an out-of-plane magnetic field component $B_z = B\cos\theta$ is introduced, it  rapidly increases the Dirac band gap $2|\Delta^{*}|$. As a result, the fixed chemical potential quickly falls within the band gap and contributions from either band are suppressed as indicated by the black arrows in Fig.~\ref{fig:nernst}(f). The sign change of the Nernst signal at $\theta_c$ stems from the reversal in sign of the Berry curvature peak as a result of the topological transition that changes the sign of the mass term $\Delta^*$. This change of sign in $\alpha_{xy}$, occurring at different polar angles $\theta$ for different azimuthal angles $\phi$ due to the phase transition boundary being dictated by $\theta_c$ (see: Eq.~(\ref{eq:critang})), is further highlighted in the full angular variation of the anomalous Nernst conductivity in Fig.~\ref{fig:nernst3D}. 

\section{\label{sec: conclusions} Conclusions}
 In summary, we discuss the emergence of a nearly half-quantized anomalous Hall and anomalous thermal Hall phase in 2D $d_{x^2-y^2}$-wave altermagnets with substrate-induced Rashba spin-orbit coupling as a result of symmetry breaking through an applied in-plane Zeeman field. The in-plane field breaks the combined fourfold rotation and time-reversal symmetry $\hat{\mathcal{C}}_{4z} \hat{\mathcal{T}}$ of the altermagnetic system and produces a gap at the Dirac point and gives rise to an effective Berry curvature monopole of $\nu = \pm 1/2$. This leads to nearly half-quantized anomalous planar Hall and anomalous planar thermal Hall responses when the chemical potential is within the Dirac gap and only one band is occupied. However, this quantization gradually diminishes either due to thermal broadening of the Fermi–Dirac distribution at elevated temperatures, which activates contributions from both bands carrying opposite Berry curvature, or by adjusting the magnetic field such that the chemical potential shifts outside the Dirac gap. Additionally, the anomalous planar Nernst conductivity peaks just outside the Zeeman gap and vanishes when the chemical potential sits in its center, with the peak position shifting outward with an increase in temperature.

 Furthermore, we show the angular dependence of all three planar responses: electrical Hall, thermal Hall, and Nernst effect. We find that these responses exhibit $\cos2\phi$ modulation inherited by the induced Dirac gap. In particular, in the presence of an in-plane magnetic field, gap closings occur \textcolor{black}{when $\hat{\mathcal{M}}_{x=\pm y}$ symmetry is restored} at $\phi=(2n+1)\pi/4$, signaling topological transitions between $\nu=+1/2$ and $\nu=-1/2$, and are manifested as sign reversals in the predicted conductivities. In the presence of an out-of-plane component of the magnetic field, the topological transitions occur at a critical value of the polar angle uniquely determined by the azimuthal angle.
 
 We propose various low-temperature experiments at accessible magnetic field strengths and analytically predict the profiles of anomalous planar Hall, anomalous planar thermal \textcolor{black}{Hall} and anomalous planar Nernst conductivities. By systematically mapping out the temperature and field dependence of these transport coefficients, our quantitative results provide clear benchmarks for experiments: one can directly compare measured transverse conductivities to the predicted plateaus and sign changes, thereby identifying the half-quantized regime in real materials.  
We also identify the regimes where the Mott relation and Wiedemann–Franz law are valid—and where they break down, offering an experimental guide to distinguish pure Berry-curvature effects from conventional quasiparticle contributions. Overall, our work reveals that planar half-quantized thermal Hall conductance plateau can occur even in systems without topological Majorana edge modes. This insight not only cautions against interpreting a half-quantized anomalous planar thermal Hall response as an unambiguous signature of chiral Majorana modes, but also highlights altermagnets as a versatile platform for realizing and detecting quantized transverse transport in zero-magnetization magnets. These findings enrich the understanding of Berry-phase-driven phenomena in altermagnets and offer new directions for exploiting their transport signatures in spintronics applications.

\section{\label{sec: Acknowledgements}Acknowledgments} S.K. and S.T. acknowledge support from SC-Quantum, ARO Grant No. W911NF2210247 and ONR Grant No. N00014-23-1-2061. 

\bibliography{quantizedTH} 

\end{document}